
\catcode`\@=11
\font\tensmc=cmcsc10      
\def\smc{\tensmc}

\def\hcorrection#1{\advance\hoffset by #1 }
\def\vcorrection#1{\advance\voffset by #1 }
\def\wlog#1{}
\newif\iftitle@
\outer\def\title{\title@true\vglue 24\p@ plus 12\p@ minus 12\p@
   \bgroup\let\\=\cr\tabskip\centering
   \halign to \hsize\bgroup\tenbf\hfill\ignorespaces##\unskip\hfill\cr}
\def\endtitle{\cr\egroup\egroup\vglue 18\p@ plus 12\p@ minus 6\p@}
\outer\def\author{\iftitle@\vglue -18\p@ plus -12\p@ minus -6\p@\fi\vglue
    12\p@ plus 6\p@ minus 3\p@\bgroup\let\\=\cr\tabskip\centering
    \halign to \hsize\bgroup\smc\hfill\ignorespaces##\unskip\hfill\cr}
\def\endauthor{\cr\egroup\egroup\vglue 18\p@ plus 12\p@ minus 6\p@}
\outer\def\heading{\bigbreak\bgroup\let\\=\cr\tabskip\centering
    \halign to \hsize\bgroup\smc\hfill\ignorespaces##\unskip\hfill\cr}
\def\endheading{\cr\egroup\egroup\nobreak\medskip}
\outer\def\subheading#1{\medbreak\noindent{\tenbf\ignorespaces
      #1\unskip.\enspace}\ignorespaces}

\outer\def\endproclaim{\par\ifdim\lastskip<\medskipamount\removelastskip
  \penalty 55 \fi\medskip\rm}
\outer\def\demo#1{\par\ifdim\lastskip<\smallskipamount\removelastskip
    \smallskip\fi\noindent{\smc\ignorespaces#1\unskip:\enspace}\rm
      \ignorespaces}

\newcount\footmarkcount@
\footmarkcount@=1
\def\makefootnote@#1#2{\insert\footins{\interlinepenalty=100
  \splittopskip=\ht\strutbox \splitmaxdepth=\dp\strutbox 
  \floatingpenalty=\@MM
  \leftskip=\z@\rightskip=\z@\spaceskip=\z@\xspaceskip=\z@
  \noindent{#1}\footstrut\rm\ignorespaces #2\strut}}
\def\footnote{\let\@sf=\empty\ifhmode\edef\@sf{\spacefactor
   =\the\spacefactor}\/\fi\futurelet\next\footnote@}
\def\footnote@{\ifx"\next\let\next\footnote@@\else
    \let\next\footnote@@@\fi\next}
\def\footnote@@"#1"#2{#1\@sf\relax\makefootnote@{#1}{#2}}
\def\footnote@@@#1{$^{\number\footmarkcount@}$\makefootnote@
   {$^{\number\footmarkcount@}$}{#1}\global\advance\footmarkcount@ by 1 }

\hyphenation{man-u-script man-u-scripts ap-pen-dix ap-pen-di-ces}
\hyphenation{data-base data-bases}
\ifx\amstexloaded@\relax\catcode`\@=13 
  \endinput\else\let\amstexloaded@=\relax\fi
\newlinechar=`\^^J
\def\eat@#1{}
\def\Space@.{\futurelet\Space@\relax}
\Space@. %
\newhelp\athelp@
{Only certain combinations beginning with @ make sense to me.^^J
Perhaps you wanted \string\@\space for a printed @?^^J
I've ignored the character or group after @.}
\def\futureletnextat@{\futurelet\next\at@}
{\catcode`\@=\active
\lccode`\Z=`\@ \lowercase
{\gdef@{\expandafter\csname futureletnextatZ\endcsname}
\expandafter\gdef\csname atZ\endcsname
   {\ifcat\noexpand\next a\def\next{\csname atZZ\endcsname}\else
   \ifcat\noexpand\next0\def\next{\csname atZZ\endcsname}\else
    \def\next{\csname atZZZ\endcsname}\fi\fi\next}
\expandafter\gdef\csname atZZ\endcsname#1{\expandafter
   \ifx\csname #1Zat\endcsname\relax\def\next
     {\errhelp\expandafter=\csname athelpZ\endcsname
      \errmessage{Invalid use of \string@}}\else
       \def\next{\csname #1Zat\endcsname}\fi\next}
\expandafter\gdef\csname atZZZ\endcsname#1{\errhelp
    \expandafter=\csname athelpZ\endcsname
      \errmessage{Invalid use of \string@}}}}
\def\atdef@#1{\expandafter\def\csname #1@at\endcsname}
\newhelp\defahelp@{If you typed \string\define\space cs instead of
\string\define\string\cs\space^^J
I've substituted an inaccessible control sequence so that your^^J
definition will be completed without mixing me up too badly.^^J
If you typed \string\define{\string\cs} the inaccessible control sequence^^J
was defined to be \string\cs, and the rest of your^^J
definition appears as input.}
\newhelp\defbhelp@{I've ignored your definition, because it might^^J
conflict with other uses that are important to me.}
\def\define{\futurelet\next\define@}
\def\define@{\ifcat\noexpand\next\relax
  \def\next{\define@@}%
  \else\errhelp=\defahelp@
  \errmessage{\string\define\space must be followed by a control 
     sequence}\def\next{\def\garbage@}\fi\next}
\def\undefined@{}
\def\preloaded@{}    
\def\define@@#1{\ifx#1\relax\errhelp=\defbhelp@
   \errmessage{\string#1\space is already defined}\def\next{\def\garbage@}%
   \else\expandafter\ifx\csname\expandafter\eat@\string
         #1@\endcsname\undefined@\errhelp=\defbhelp@
   \errmessage{\string#1\space can't be defined}\def\next{\def\garbage@}%
   \else\expandafter\ifx\csname\expandafter\eat@\string#1\endcsname\relax
     \def\next{\def#1}\else\errhelp=\defbhelp@
     \errmessage{\string#1\space is already defined}\def\next{\def\garbage@}%
      \fi\fi\fi\next}
\def\famzero{\fam\z@}

\def\lim{\mathop{\famzero lim}}

\def\textfont@#1#2{\def#1{\relax\ifmmode
    \errmessage{Use \string#1\space only in text}\else#2\fi}}
\textfont@\rm\tenrm
\textfont@\it\tenit
\textfont@\sl\tensl
\textfont@\bf\tenbf
\textfont@\smc\tensmc
\let\ic@=\/
\def\/{\unskip\ic@}
\def\textfonti{\the\textfont1 }
\def\t#1#2{{\edef\next{\the\font}\textfonti\accent"7F \next#1#2}}
\let\B=\=
\let\D=\.
\def~{\unskip\nobreak\ \ignorespaces}
{\catcode`\@=\active
\gdef\@{\char'100 }}
\atdef@-{\leavevmode\futurelet\next\athyph@}
\def\athyph@{\ifx\next-\let\next=\athyph@@
  \else\let\next=\athyph@@@\fi\next}
\def\athyph@@@{\hbox{-}}
\def\athyph@@#1{\futurelet\next\athyph@@@@}
\def\athyph@@@@{\if\next-\def\next##1{\hbox{---}}\else
    \def\next{\hbox{--}}\fi\next}
\def\.{.\spacefactor=\@m}
\atdef@.{\null.}
\atdef@,{\null,}
\atdef@;{\null;}
\atdef@:{\null:}
\atdef@?{\null?}
\atdef@!{\null!}   
\def\srdr@{\thinspace}                     
\def\drsr@{\kern.02778em}
\def\sldl@{\kern.02778em}
\def\dlsl@{\thinspace}
\atdef@"{\unskip\futurelet\next\atqq@}
\def\atqq@{\ifx\next\Space@\def\next. {\atqq@@}\else
         \def\next.{\atqq@@}\fi\next.}
\def\atqq@@{\futurelet\next\atqq@@@}
\def\atqq@@@{\ifx\next`\def\next`{\atqql@}\else\def\next'{\atqqr@}\fi\next}
\def\atqql@{\futurelet\next\atqql@@}
\def\atqql@@{\ifx\next`\def\next`{\sldl@``}\else\def\next{\dlsl@`}\fi\next}
\def\atqqr@{\futurelet\next\atqqr@@}
\def\atqqr@@{\ifx\next'\def\next'{\srdr@''}\else\def\next{\drsr@'}\fi\next}

\def\textfontii{\the\textfont2 }
\def\{{\relax\ifmmode\lbrace\else
    {\textfontii f}\spacefactor=\@m\fi}
\def\}{\relax\ifmmode\rbrace\else
    \let\@sf=\empty\ifhmode\edef\@sf{\spacefactor=\the\spacefactor}\fi
      {\textfontii g}\@sf\relax\fi}   
\def\nonhmodeerr@#1{\errmessage
     {\string#1\space allowed only within text}}
\def\linebreak{\relax\ifhmode\unskip\break\else
    \nonhmodeerr@\linebreak\fi}
\def\allowlinebreak{\relax
   \ifhmode\allowbreak\else\nonhmodeerr@\allowlinebreak\fi}
\newskip\saveskip@
\def\nolinebreak{\relax\ifhmode\saveskip@=\lastskip\unskip
  \nobreak\ifdim\saveskip@>\z@\hskip\saveskip@\fi
   \else\nonhmodeerr@\nolinebreak\fi}
\def\newline{\relax\ifhmode\null\hfil\break
    \else\nonhmodeerr@\newline\fi}
\def\nonmathaerr@#1{\errmessage
     {\string#1\space is not allowed in display math mode}}
\def\nonmathberr@#1{\errmessage{\string#1\space is allowed only in math mode}}
\def\mathbreak{\relax\ifmmode\ifinner\break\else
   \nonmathaerr@\mathbreak\fi\else\nonmathberr@\mathbreak\fi}
\def\nomathbreak{\relax\ifmmode\ifinner\nobreak\else
    \nonmathaerr@\nomathbreak\fi\else\nonmathberr@\nomathbreak\fi}
\def\allowmathbreak{\relax\ifmmode\ifinner\allowbreak\else
     \nonmathaerr@\allowmathbreak\fi\else\nonmathberr@\allowmathbreak\fi}
\def\pagebreak{\relax\ifmmode
   \ifinner\errmessage{\string\pagebreak\space
     not allowed in non-display math mode}\else\postdisplaypenalty-\@M\fi
   \else\ifvmode\penalty-\@M\else\edef\spacefactor@
       {\spacefactor=\the\spacefactor}\vadjust{\penalty-\@M}\spacefactor@
        \relax\fi\fi}
\def\nopagebreak{\relax\ifmmode
     \ifinner\errmessage{\string\nopagebreak\space
    not allowed in non-display math mode}\else\postdisplaypenalty\@M\fi
    \else\ifvmode\nobreak\else\edef\spacefactor@
        {\spacefactor=\the\spacefactor}\vadjust{\penalty\@M}\spacefactor@
         \relax\fi\fi}
\def\newpage{\relax\ifvmode\vfill\penalty-\@M\else\nonvmodeerr@\newpage\fi}
\def\nonvmodeerr@#1{\errmessage
    {\string#1\space is allowed only between paragraphs}}
\def\smallpagebreak{\relax\ifvmode\smallbreak
      \else\nonvmodeerr@\smallpagebreak\fi}
\def\medpagebreak{\relax\ifvmode\medbreak
       \else\nonvmodeerr@\medpagebreak\fi}
\def\bigpagebreak{\relax\ifvmode\bigbreak
      \else\nonvmodeerr@\bigpagebreak\fi}
\newdimen\captionwidth@
\captionwidth@=\hsize
\advance\captionwidth@ by -1.5in
\def\caption#1{}
\def\topspace#1{\gdef\thespace@{#1}\ifvmode\def\next
    {\futurelet\next\topspace@}\else\def\next{\nonvmodeerr@\topspace}\fi\next}
\def\topspace@{\ifx\next\Space@\def\next. {\futurelet\next\topspace@@}\else
     \def\next.{\futurelet\next\topspace@@}\fi\next.}
\def\topspace@@{\ifx\next\caption\let\next\topspace@@@\else
    \let\next\topspace@@@@\fi\next}
 \def\topspace@@@@{\topinsert\vbox to 
       \thespace@{}\endinsert}
\def\topspace@@@\caption#1{\topinsert\vbox to
    \thespace@{}\nobreak
      \smallskip
    \setbox\z@=\hbox{\noindent\ignorespaces#1\unskip}%
   \ifdim\wd\z@>\captionwidth@
   \centerline{\vbox{\hsize=\captionwidth@\noindent\ignorespaces#1\unskip}}%
   \else\centerline{\box\z@}\fi\endinsert}
\def\midspace#1{\gdef\thespace@{#1}\ifvmode\def\next
    {\futurelet\next\midspace@}\else\def\next{\nonvmodeerr@\midspace}\fi\next}
\def\midspace@{\ifx\next\Space@\def\next. {\futurelet\next\midspace@@}\else
     \def\next.{\futurelet\next\midspace@@}\fi\next.}
\def\midspace@@{\ifx\next\caption\let\next\midspace@@@\else
    \let\next\midspace@@@@\fi\next}
 \def\midspace@@@@{\midinsert\vbox to 
       \thespace@{}\endinsert}
\def\midspace@@@\caption#1{\midinsert\vbox to
    \thespace@{}\nobreak
      \smallskip
      \setbox\z@=\hbox{\noindent\ignorespaces#1\unskip}%
      \ifdim\wd\z@>\captionwidth@
    \centerline{\vbox{\hsize=\captionwidth@\noindent\ignorespaces#1\unskip}}%
    \else\centerline{\box\z@}\fi\endinsert}
\mathchardef\prime@="0230
\def\prime{{{}\prime@{}}}
\def\prim@s{\prime@\futurelet\next\pr@m@s}

\def\,{\relax\ifmmode\mskip\thinmuskip\else\thinspace\fi}
\def\!{\relax\ifmmode\mskip-\thinmuskip\else\negthinspace\fi}
\def\frac#1#2{{#1\over#2}}

\def\:{\nobreak\hskip.1111em{:}\hskip.3333em plus .0555em\relax}
\def\intic@{\mathchoice{\hskip5\p@}{\hskip4\p@}{\hskip4\p@}{\hskip4\p@}}
\def\negintic@
 {\mathchoice{\hskip-5\p@}{\hskip-4\p@}{\hskip-4\p@}{\hskip-4\p@}}
\def\intkern@{\mathchoice{\!\!\!}{\!\!}{\!\!}{\!\!}}
\def\intdots@{\mathchoice{\cdots}{{\cdotp}\mkern1.5mu
    {\cdotp}\mkern1.5mu{\cdotp}}{{\cdotp}\mkern1mu{\cdotp}\mkern1mu
      {\cdotp}}{{\cdotp}\mkern1mu{\cdotp}\mkern1mu{\cdotp}}}
\newcount\intno@             
\def\iint{\intno@=\tw@\futurelet\next\ints@} 
\def\iiint{\intno@=\thr@@\futurelet\next\ints@}
\def\iiiint{\intno@=4 \futurelet\next\ints@}
\def\idotsint{\intno@=\z@\futurelet\next\ints@}
\def\ints@{\findlimits@\ints@@}
\newif\iflimtoken@
\newif\iflimits@
\def\findlimits@{\limtoken@false\limits@false\ifx\next\limits
 \limtoken@true\limits@true\else\ifx\next\nolimits\limtoken@true\limits@false
    \fi\fi}
\def\multintlimits@{\intop\ifnum\intno@=\z@\intdots@
  \else\intkern@\fi
    \ifnum\intno@>\tw@\intop\intkern@\fi
     \ifnum\intno@>\thr@@\intop\intkern@\fi\intop}
\def\multint@{\int\ifnum\intno@=\z@\intdots@\else\intkern@\fi
   \ifnum\intno@>\tw@\int\intkern@\fi
    \ifnum\intno@>\thr@@\int\intkern@\fi\int}
\def\ints@@{\iflimtoken@\def\ints@@@{\iflimits@
   \negintic@\mathop{\intic@\multintlimits@}\limits\else
    \multint@\nolimits\fi\eat@}\else
     \def\ints@@@{\multint@\nolimits}\fi\ints@@@}
\def\Sb{_\bgroup\vspace@
        \baselineskip=\fontdimen10 \scriptfont\tw@
        \advance\baselineskip by \fontdimen12 \scriptfont\tw@
        \lineskip=\thr@@\fontdimen8 \scriptfont\thr@@
        \lineskiplimit=\thr@@\fontdimen8 \scriptfont\thr@@
        \Let@\vbox\bgroup\halign\bgroup \hfil$\scriptstyle
            {##}$\hfil\cr}
\def\endSb{\crcr\egroup\egroup\egroup}
\def\Sp{^\bgroup\vspace@
        \baselineskip=\fontdimen10 \scriptfont\tw@
        \advance\baselineskip by \fontdimen12 \scriptfont\tw@
        \lineskip=\thr@@\fontdimen8 \scriptfont\thr@@
        \lineskiplimit=\thr@@\fontdimen8 \scriptfont\thr@@
        \Let@\vbox\bgroup\halign\bgroup \hfil$\scriptstyle
            {##}$\hfil\cr}
\def\endSp{\crcr\egroup\egroup\egroup}
\def\Let@{\relax\iffalse{\fi\let\\=\cr\iffalse}\fi}
\def\vspace@{\def\vspace##1{\noalign{\vskip##1 }}}
\def\aligned{\,\vcenter\bgroup\vspace@\Let@\openup\jot\m@th\ialign
  \bgroup \strut\hfil$\displaystyle{##}$&$\displaystyle{{}##}$\hfil\crcr}
\def\endaligned{\crcr\egroup\egroup}
\def\matrix{\,\vcenter\bgroup\Let@\vspace@
    \normalbaselines
  \m@th\ialign\bgroup\hfil$##$\hfil&&\quad\hfil$##$\hfil\crcr
    \mathstrut\crcr\noalign{\kern-\baselineskip}}
\def\endmatrix{\crcr\mathstrut\crcr\noalign{\kern-\baselineskip}\egroup
                \egroup\,}
\newtoks\hashtoks@
\hashtoks@={#}
\def\format{\crcr\egroup\iffalse{\fi\ifnum`}=0 \fi\format@}
\def\format@#1\\{\def\preamble@{#1}%
  \def\c{\hfil$\the\hashtoks@$\hfil}%
  \def\r{\hfil$\the\hashtoks@$}%
  \def\l{$\the\hashtoks@$\hfil}%
  \setbox\z@=\hbox{\xdef\Preamble@{\preamble@}}\ifnum`{=0 \fi\iffalse}\fi
   \ialign\bgroup\span\Preamble@\crcr}

\def\cases{\left\{\,\vcenter\bgroup\vspace@
     \normalbaselines\openup\jot\m@th
       \Let@\ialign\bgroup$##$\hfil&\quad$##$\hfil\crcr
      \mathstrut\crcr\noalign{\kern-\baselineskip}}

\newif\iftagsleft@
\tagsleft@true
\def\TagsOnRight{\global\tagsleft@false}
\def\tag#1$${\iftagsleft@\leqno\else\eqno\fi
 \hbox{\def\pagebreak{\global\postdisplaypenalty-\@M}%
 \def\nopagebreak{\global\postdisplaypenalty\@M}\rm(#1\unskip)}%
  $$\postdisplaypenalty\z@\ignorespaces}
\interdisplaylinepenalty=\@M
\def\allowdisplaybreak@{\def\allowdisplaybreak{\noalign{\allowbreak}}}
\def\displaybreak@{\def\displaybreak{\noalign{\break}}}
\def\align#1\endalign{\def\tag{&}\vspace@\allowdisplaybreak@\displaybreak@
  \iftagsleft@\lalign@#1\endalign\else
   \ralign@#1\endalign\fi}
\def\ralign@#1\endalign{\displ@y\Let@\tabskip\centering\halign to\displaywidth
     {\hfil$\displaystyle{##}$\tabskip=\z@&$\displaystyle{{}##}$\hfil
       \tabskip=\centering&\llap{\hbox{(\rm##\unskip)}}\tabskip\z@\crcr
             #1\crcr}}
\def\lalign@
 #1\endalign{\displ@y\Let@\tabskip\centering\halign to \displaywidth
   {\hfil$\displaystyle{##}$\tabskip=\z@&$\displaystyle{{}##}$\hfil
   \tabskip=\centering&\kern-\displaywidth
        \rlap{\hbox{(\rm##\unskip)}}\tabskip=\displaywidth\crcr
               #1\crcr}}
\def\overrightarrow{\mathpalette\overrightarrow@}
\def\overrightarrow@#1#2{\vbox{\ialign{$##$\cr
    #1{-}\mkern-6mu\cleaders\hbox{$#1\mkern-2mu{-}\mkern-2mu$}\hfill
     \mkern-6mu{\to}\cr
     \noalign{\kern -1\p@\nointerlineskip}
     \hfil#1#2\hfil\cr}}}
\def\overleftarrow{\mathpalette\overleftarrow@}
\def\overleftarrow@#1#2{\vbox{\ialign{$##$\cr
     #1{\leftarrow}\mkern-6mu\cleaders\hbox{$#1\mkern-2mu{-}\mkern-2mu$}\hfill
      \mkern-6mu{-}\cr
     \noalign{\kern -1\p@\nointerlineskip}
     \hfil#1#2\hfil\cr}}}
\def\overleftrightarrow{\mathpalette\overleftrightarrow@}
\def\overleftrightarrow@#1#2{\vbox{\ialign{$##$\cr
     #1{\leftarrow}\mkern-6mu\cleaders\hbox{$#1\mkern-2mu{-}\mkern-2mu$}\hfill
       \mkern-6mu{\to}\cr
    \noalign{\kern -1\p@\nointerlineskip}
      \hfil#1#2\hfil\cr}}}
\def\underrightarrow{\mathpalette\underrightarrow@}
\def\underrightarrow@#1#2{\vtop{\ialign{$##$\cr
    \hfil#1#2\hfil\cr
     \noalign{\kern -1\p@\nointerlineskip}
    #1{-}\mkern-6mu\cleaders\hbox{$#1\mkern-2mu{-}\mkern-2mu$}\hfill
     \mkern-6mu{\to}\cr}}}
\def\underleftarrow{\mathpalette\underleftarrow@}
\def\underleftarrow@#1#2{\vtop{\ialign{$##$\cr
     \hfil#1#2\hfil\cr
     \noalign{\kern -1\p@\nointerlineskip}
     #1{\leftarrow}\mkern-6mu\cleaders\hbox{$#1\mkern-2mu{-}\mkern-2mu$}\hfill
      \mkern-6mu{-}\cr}}}
\def\underleftrightarrow{\mathpalette\underleftrightarrow@}
\def\underleftrightarrow@#1#2{\vtop{\ialign{$##$\cr
      \hfil#1#2\hfil\cr
    \noalign{\kern -1\p@\nointerlineskip}
     #1{\leftarrow}\mkern-6mu\cleaders\hbox{$#1\mkern-2mu{-}\mkern-2mu$}\hfill
       \mkern-6mu{\to}\cr}}}
\def\sqrt#1{\radical"270370 {#1}}
\def\dots{\relax\ifmmode\let\next=\ldots\else\let\next=\tdots@\fi\next}
\def\tdots@{\unskip\ \tdots@@}
\def\tdots@@{\futurelet\next\tdots@@@}
\def\tdots@@@{$\mathinner{\ldotp\ldotp\ldotp}\,
   \ifx\next,$\else
   \ifx\next.\,$\else
   \ifx\next;\,$\else
   \ifx\next:\,$\else
   \ifx\next?\,$\else
   \ifx\next!\,$\else
   $ \fi\fi\fi\fi\fi\fi}
\def\text{\relax\ifmmode\let\next=\text@\else\let\next=\text@@\fi\next}
\def\text@@#1{\hbox{#1}}
\def\text@#1{\mathchoice
 {\hbox{\everymath{\displaystyle}\def\textfonti{\the\textfont1 }%
    \def\textfontii{\the\textfont2 }\textdef@@ T#1}}
 {\hbox{\everymath{\textstyle}\def\textfonti{\the\textfont1 }%
    \def\textfontii{\the\textfont2 }\textdef@@ T#1}}
 {\hbox{\everymath{\scriptstyle}\def\textfonti{\the\scriptfont1 }%
   \def\textfontii{\the\scriptfont2 }\textdef@@ S\rm#1}}
 {\hbox{\everymath{\scriptscriptstyle}\def\textfonti{\the\scriptscriptfont1 }%
   \def\textfontii{\the\scriptscriptfont2 }\textdef@@ s\rm#1}}}
\def\textdef@@#1{\textdef@#1\rm \textdef@#1\bf
   \textdef@#1\sl \textdef@#1\it}

\def\textdef@#1#2{\def\next{\csname\expandafter\eat@\string#2fam\endcsname}%
\if S#1\edef#2{\the\scriptfont\next\relax}%
 \else\if s#1\edef#2{\the\scriptscriptfont\next\relax}%
 \else\edef#2{\the\textfont\next\relax}\fi\fi}
\scriptfont\itfam=\tenit \scriptscriptfont\itfam=\tenit
\scriptfont\slfam=\tensl \scriptscriptfont\slfam=\tensl
\mathcode`\0="0030
\mathcode`\1="0031
\mathcode`\2="0032
\mathcode`\3="0033
\mathcode`\4="0034
\mathcode`\5="0035
\mathcode`\6="0036
\mathcode`\7="0037
\mathcode`\8="0038
\mathcode`\9="0039
\def\Cal{\relax\ifmmode\let\next=\Cal@\else
     \def\next{\errmessage{Use \string\Cal\space only in math mode}}\fi\next}
\def\Cal@#1{{\fam2 #1}}
\def\bold{\relax\ifmmode\let\next=\bold@\else
   \def\next{\errmessage{Use \string\bold\space only in math
      mode}}\fi\next}\def\bold@#1{{\fam\bffam #1}}
\mathchardef\Gamma="0000
\mathchardef\Delta="0001
\mathchardef\Theta="0002
\mathchardef\Lambda="0003
\mathchardef\Xi="0004
\mathchardef\Pi="0005
\mathchardef\Sigma="0006
\mathchardef\Upsilon="0007
\mathchardef\Phi="0008
\mathchardef\Psi="0009
\mathchardef\Omega="000A
\mathchardef\varGamma="0100
\mathchardef\varDelta="0101
\mathchardef\varTheta="0102
\mathchardef\varLambda="0103
\mathchardef\varXi="0104
\mathchardef\varPi="0105
\mathchardef\varSigma="0106
\mathchardef\varUpsilon="0107
\mathchardef\varPhi="0108
\mathchardef\varPsi="0109
\mathchardef\varOmega="010A
\font\dummyft@=dummy
\fontdimen1 \dummyft@=\z@
\fontdimen2 \dummyft@=\z@
\fontdimen3 \dummyft@=\z@
\fontdimen4 \dummyft@=\z@
\fontdimen5 \dummyft@=\z@
\fontdimen6 \dummyft@=\z@
\fontdimen7 \dummyft@=\z@
\fontdimen8 \dummyft@=\z@
\fontdimen9 \dummyft@=\z@
\fontdimen10 \dummyft@=\z@
\fontdimen11 \dummyft@=\z@
\fontdimen12 \dummyft@=\z@
\fontdimen13 \dummyft@=\z@
\fontdimen14 \dummyft@=\z@
\fontdimen15 \dummyft@=\z@
\fontdimen16 \dummyft@=\z@
\fontdimen17 \dummyft@=\z@
\fontdimen18 \dummyft@=\z@
\fontdimen19 \dummyft@=\z@
\fontdimen20 \dummyft@=\z@
\fontdimen21 \dummyft@=\z@
\fontdimen22 \dummyft@=\z@
\def\fontlist@{\\{\tenrm}\\{\sevenrm}\\{\fiverm}\\{\teni}\\{\seveni}%
 \\{\fivei}\\{\tensy}\\{\sevensy}\\{\fivesy}\\{\tenex}\\{\tenbf}\\{\sevenbf}%
 \\{\fivebf}\\{\tensl}\\{\tenit}\\{\tensmc}}
\def\dodummy@{{\def\\##1{\global\let##1=\dummyft@}\fontlist@}}
\newif\ifsyntax@
\newcount\countxviii@
\def\newtoks@{\alloc@5\toks\toksdef\@cclvi}
\def\nopages@{\output={\setbox\z@=\box\@cclv \deadcycles=\z@}\newtoks@\output}
\def\syntax{\syntax@true\dodummy@\countxviii@=\count18
\loop \ifnum\countxviii@ > \z@ \textfont\countxviii@=\dummyft@
   \scriptfont\countxviii@=\dummyft@ \scriptscriptfont\countxviii@=\dummyft@
     \advance\countxviii@ by-\@ne\repeat
\dummyft@\tracinglostchars=\z@
  \nopages@\frenchspacing\hbadness=\@M}
\def\magstep#1{\ifcase#1 1000\or
 1200\or 1440\or 1728\or 2074\or 2488\or 
 \errmessage{\string\magstep\space only works up to 5}\fi\relax}
{\lccode`\2=`\p \lccode`\3=`\t 
 \lowercase{\gdef\tru@#123{#1truept}}}

\def\scaletype#1{\mag=#1\relax
 \hsize=\expandafter\tru@\the\hsize
 \vsize=\expandafter\tru@\the\vsize
 \dimen\footins=\expandafter\tru@\the\dimen\footins}

\def\scalefont#1#2\andcallit#3{\edef\font@{\the\font}#1\font#3=
  \fontname\font\space scaled #2\relax\font@}
\def\Mag@#1#2{\ifdim#1<1pt\multiply#1 #2\relax\divide#1 1000 \else
  \ifdim#1<10pt\divide#1 10 \multiply#1 #2\relax\divide#1 100\else
  \divide#1 100 \multiply#1 #2\relax\divide#1 10 \fi\fi}
\def\scalelinespacing#1{\Mag@\baselineskip{#1}\Mag@\lineskip{#1}%
  \Mag@\lineskiplimit{#1}}
\def\wlog#1{\immediate\write-1{#1}}
\catcode`\@=\active


\input vanilla.sty


\magnification 1200 
\hsize=16.6truecm
\vsize=22.6truecm 

\font\eightrm=cmr8

\font\eightsl=cmsl8

\font\gross=cmcsc10

 1

\font\tit=cmti10 scaled\magstep 2      
 2

\nopagenumbers
\footline={\hss}

\TagsOnRight

\csname lrh\endcsname
\csname rrh\endcsname
\newcount\titel
\headline={\ifnum\titel=0\eightrm\ifodd\pageno\hss
\rrh\qquad\ \folio\else\folio
            \qquad\ \lrh
            \hss\fi\else%
             \hss\global\titel=0\fi}


\hyphenation{
Meas-ure-ment Meas-uring 
meas-ure-ment meas-ure meas-uring
pre-meas-ure-ment pre-meas-ure pre-meas-uring}


\def\qm{quantum mechanics}
                       
\def\me{meas\-ure}                    \def\mt{\me{}ment} \def\mts{\mt{}s}

\def\op{operator}                    
\def\ob{observable}                  
\def\sad{self-adjoint}               \def\sop{\sad\ \op}

\def\bo#1{{\bold #1}}                    

\def\no#1{\left\|#1\right\|}             
\def\ip#1#2{\left\langle\,#1\,|\,#2\,\right\rangle} 
\def\kb#1#2{|#1\,\rangle\langle\,#2|}    
\def\ket#1{\mid#1\rangle}                
           
    \def\ts12{{\textstyle{\frac 12}}}


\def\fii{\varphi}   
                       \def\cs{$\Cal S$}








                 \def\12pi{\frac 1{2\pi}}


\overfullrule 0pt
\advance\hoffset by -1cm

\baselineskip 18pt

\parskip 0pt plus 1pt

\font\gross=cmcsc10
 2
 1
\font\eightrm=cmr8
\font\eightsl=cmsl8

\font\tit=cmti10 scaled \magstep 2

\def\R{\hbox{I\kern -0.2em R}}
\def\1{\hbox{1\kern -0.3em I}}

\def\no1#1{\|#1\|_1}

\def\isd{{individual state determination}}
\def\soh{spin-$\frac 12$}

\headline={\ifnum\titel=0\eightrm
  P.\ Busch\quad -- \quad Is the Quantum State (an) Observable?
        \hss \folio
            \else%
             \hss\global\titel=0\fi}

\newcount\titel
\titel=1
\footline={\hss}

\title{\tit Is the quantum state (an) observable?\footnote{
{\eightrm In:} {\eightsl Experimental Metaphysics -- Quantum Mechanical
Studies in Honor of Abner Shimony.} \newline
{\eightrm Eds.\ R.S.\ Cohen and J.\
Stachel. D.\ Reidel, Dordrecht, to be published, 1996.}}}
\endtitle

\vskip .2truecm
\centerline{\gross P.\ Busch}

\centerline{
Department of Applied Mathematics, The University of Hull}
\centerline{Kingston upon Hull HU6 7RX, United Kingdom}
\centerline{Electronic address: p.busch\@{}maths.hull.ac.uk}
\vskip .5truecm

\centerline{\gross Abstract}
\vskip 3pt

{\advance\baselineskip by -6pt
{\narrower{\eightrm
\noindent
 We explore the sense in which the state of a physical
system may or may not be regarded (an) observable in quantum mechanics.
Simple and general arguments from various lines of approach
are reviewed which demonstrate the following no-go claims:
(1) the structure of quantum mechanics precludes the
determination of the state of a single system by means of
meas\-urements performed on that system only; (2) there is
no way of using entangled two-particle states to transmit
superluminal signals. Employing the representation of
observables as general positive operator valued measures,
our analysis allows one to indicate whether optimal separation
of different states  is achieved by means of sharp or
unsharp observables.

}}}


\subheading{1.\ Introduction}

\noindent
 Quantum mechanics is often claimed to be a theory about
ensembles only rather than about single systems. Yet there
is an increasing variety of experiments exhibiting
individual quantum processes which were conceived, devised
and explained on the basis of this very theory. Therefore,
in order to reach a proper appreciation of the scope of
quantum mechanics, it is necessary to spell out the senses
in which the theory does or does not apply to individual
systems.  In this contribution we begin with highlighting
the individual aspects of quantum mechanics (Section 2) and
proceed then to show why, in particular, the determination
of the state of an individual system ought to be impossible
for various reasons of consistency (Section 3). Rigorous
arguments demonstrating this impossibility are then
reviewed, using the general representation of observables as
positive operator valued measures (Section 4). Finally we
address briefly the question as to how well two
non-orthogonal states can be discriminated (Section 5).

\subheading{2.\ Individual aspects of quantum mechanics}

\noindent
 Quantum mechanics is commonly formulated in terms of the
basic duality of states and observables. Given a pair of a
state operator $\rho$ and a \sop\ $A$ representing a
physical quantity, the number $\text{Trace}[\rho A]$ is
interpreted as the expectation value of the quantity if
measured on an ensemble of systems all prepared in the state
described by $\rho$. Using the spectral decomposition of
$A$, this minimal statistical interpretation extends to the
probability distribution $p_\rho^A$ determined by $\rho$ and
$A$, of which the expectation value is the first moment.

In this way the empirical meaning of certain quantum
mechanical terms is fixed by making reference to the
probability structure carried by the theory. But this very
structure is based on the lattice of subspaces, or
orthogonal projections, of the underlying Hilbert space.
This observation opens up the possibility to introduce
relations between states and observables which are
non-probabilistic in the first instance (though equivalent,
probabilistic formulations do exist): a given state $\rho$
may be an eigenstate of a projection $P$, meaning that one
of the following equations holds:
 $$
P\rho=\rho,\quad\text{ or }\quad P\rho=O.\tag 1 
$$
 In such a case a system prepared in the state $\rho$ can be
said to possess the {\it real property} represented either
by $P$ or its complement $I-P$. So $P$ is {\it real} in
$\rho$ if $P\rho=\rho$, {\it absent} in $\rho$ if $P\rho=O$,
and {\it indeterminate} in $\rho$ if it is neither real nor
absent. The quantum state is thus found to have the
undoubtedly individual aspect of defining a valuation on
the set of potential properties represented by the
projections. We spell this out with reference to a pure
state, represented by a unit vector $\psi$ of the Hilbert
space, and discrete observables.

Every observable $A$ of which $\psi$ is an eigenvector has a
definite value, and the outcome of a meas\-urement of such
an observable $A$ can be predicted with certainty to yield
the corresponding eigenvalue as an outcome. It is a
distinctive feature of quantum mechanics that for any state
there are observables of which that state is no
eigenvector. This corresponds to the situation where $\psi$
is a (proper) superposition of eigenstates of an observable
$B$, in which case one can still (or only) assert that that
observable is {\it indeterminate} in the state $\psi$, that
is, that $B$ does not have a definite value.  Considering an
ideal, repeatable meas\-urement of $B=\sum
b_k\kb{\psi_k}{\psi_k}$, then each single system of a
collection of equally prepared systems can be predicted to
{\it jump} into one of the eigenstates $\psi_k$, with the
probability $p_\psi^B(b_k)=\big|\ip{\psi_k}{\psi}
\big|^2$. The value of the observable $B$ thus {\it becomes
determined} dynamically for each system in the course of
such a measurement and can thus be ascertained to be a real
property of the system in its state after the measurement.

Quantum mechanics does also allow for the preparation of an
individual system in a state $\psi$: one may simply perform
a repeatable \mt\ of an observable of which $\psi$ is an
eigenstate (associated with a nondegenerate eigenvalue).  As
long as there is a nonzero probability for the corresponding
eigenvalue, there will be a chance to obtain the desired
state. Hence a repeatable \mt\ can be used as a filter for
preparing individual systems in known states.

So far we have elucidated the possibilities of obtaining
information about properties or state descriptions
pertaining to a single system {\it after} a \mt. The
complementary question is whether, or to what extent,
quantum mechanics allows one to infer, solely on the basis
of information obtained from a \mt\ or any sequence of
manipulations carried out on a single system, in which state
that system was {\it prior} to the \mt. This question of
{\it individual state determination} has been raised, in
particular, in the context of discussions about proposals of
employing entangled quantum states of spatially separated
systems to effect superluminal signaling [1].  For a long
time arguments against such a violation of Einstein
causality were formulated on the level of more or less
realistic concrete models of signaling schemes, including
considerations of amplification or photon cloning processes
[2-9]. More recently the issue was touched upon again in
quantum information theory when it was realized that quantum
correlations can be used for the safe teleportation of
cryptographic keys [10]; yet another context where the
question of \isd\ arises naturally is the quantum
thermodynamics of black holes [11].  Meanwhile there do
exist {\it abstract} and thus completely {\it general}
proofs of the impossibility of \isd\ and some systematic
investigations on the problem of optimizing the state
inference [11,12-15]. However, these investigations seem to
have remained largely unnoticed, and the production of new
schemes for \isd{}s, while {\it apriori} doomed to fail,
does not seem to come to an end [16]. Therefore it may be
justified to collect those simple and general arguments and
show how \qm\ manages to preclude \isd\ and to protect
itself against the otherwise desastrous implications. The
presentation will be self-contained and occasionally uses
methods that are different from those applied in the
literature.  We proceed with outlining first why \isd\ should be
regarded undesirable.

\subheading{3.\ Undesirable consequences of \isd}

\noindent
 Imagine that by some ingenious {\it ISD-procedure} it were
possible to determine the state of a quantum system,
represented by a density operator $T$. It follows, first of
all, that one were able to decide whether a given quantum
system is in a pure or a mixed state.
(Of course, it would also be possible to distinguish between
arbitrary pairs of pure states; we return to this case below.) 
This means that a
single \mt\ on the given system alone will suffice to find
out whether or not it is entangled with some environment
systems. Consequently, one would have to realize that the
usual \qm\ description of states as density operators is
{\it incomplete}: the same density operator $T$ would
represent two physically distinguishable situations: (1) an
ensemble of systems in pure states, the distribution of
which is described by $T$; or (2) an ensemble of systems,
each entangled with another system, so that in each single
case $T$ provides the exhaustive description of the {\it
reduced} state.  

Due to this feature, the {\it ISD}-procedure can be used to effect
superluminal signal transmission. In fact, consider an
Einstein-Podolsky-Rosen correlated system consisting of two
spatially separated \soh\ particles $\Cal A,\Cal B$ in a
singlet state,
 $$
\Psi\ =\ \frac 1{\sqrt 2}\big(\ket{+,z}\ket{-,z}-\ket{-,z}\ket{+,z}\big).
\tag 2
$$
 (Here $\ket{\pm,z}$ denotes the eigenvectors of the \soh\
operator $s_z$.) A signal would consist of one observer's,
{\it Bob's}, measuring or not measuring some spin component
$s_{\bo n}$ on particle $\Cal B$, while the second observer,
{\it Abner}, would apply the {\it ISD}-procedure on particle $\Cal
A$ to find out whether {\it Bob} has measured or not: in the first
case {\it Abner} would find $\Cal A$ in one of the pure states
$\ket{\pm,\bo n}$, while in the latter case he would find $\Cal A$
in the mixed reduced state $\frac
12\big(\kb{+,z}{+,z}+\kb{-,z}{-,z}\big)=\frac 12 I$.  Hence the
{\it ISD}-procedure would allow an instantaneous detection of
which bit of the message was sent from the spacelike
distant particle.

We note  that the alphabet can be easily extended: instead
of measuring or not, {\it Bob} may choose to measure one of a
collection of $N$ spin components $s_{\bo n_k}$,
$k=1,\cdots,N$.  In that case {\it Abner's} {\it ISD}-procedure would
have to be able only to distinguish between a finite number
(at least 4) of pure states. We show next that \isd\
is indeed impossible in general and also that the
unambiguous distinction even of only two non-orthogonal
states must fail.

\subheading{4.\ Impossibility of individual state determination}

\subheading{4.1} {\it State inference from \mt\ outcomes.}
 The determination of a system's state  requires that some
\mt\ is performed.  Every \mt\ can be represented by some
\ob\ whose values are exhibited as the \mt\ outcomes. In order
to decide from the outcomes whether or not the system was in
a given state $\fii$, there needs to be at least one outcome which
occurs with certainty if the state was $\fii$, and which
will certainly not occur if the state was some state $\psi$ different
from $\fii$. Hence the probability for that outcome must be
$p_\fii=1$ for the state $\fii$ and $p_\psi=0$ for $\psi$. Since the quantum
mechanical probabilities are expectations of some positive
operators $E$ representing the event in question, it follows
that $\ip\fii{E\fii}=1$ and
$\ip{\psi}{E\psi}=0$. These equations are equivalent to
$$
E\fii=\fii\quad\text{ and }\quad E\psi=O\tag 3
$$
(cf.\ remark [17]). Therefore,
 $$
\ip\fii{\psi}\ =\ \ip{E\fii}{\psi}\ =\ 
\ip{\fii}{E\psi}\ =\ 0,\tag 4
$$
 which is to say that $\fii$ and $\psi$ are mutually
orthogonal. It follows that no \mt\ exists which would allow
one to distinguish unequivocally between any (even a single)
pair of non-orthogonal states. 

It may be noted that in this argument
the usual assumption is {\it not} made that the operator $E$
is a projection. For the numbers $\ip\xi{E\xi}$ to represent
probabilities, $\xi$ being any unit vector, it is formally
necessary and sufficient that $E$ is an operator bounded
between the unit ($I$) and null ($O$) operators. These
operators are known as {\it effects} (as compared to
properties). An effect for which there are states such that
(3) is satisfied belongs to the class of {\it approximate
properties} [18]; for $E$ it can be asserted that in the state
$\fii$, $E$ is real while for $\psi$ the complement $I-E$ is
real. 

\subheading{4.2} {\it Measurement theoretical formulation.}
There is a \mt\ theoretical version of the argument which
exploits the unitarity of the quantum dynamics. This
formulation has the advantage of allowing a direct
confrontation with the concrete model proposals put forward
in favor of \isd\ [1,16]. In the quantum mechanical
description of a \mt\ process, the object system, originally
in state $\fii$, is coupled with an apparatus (or probe),
originally in state $\phi$, by means of a unitary operator
$U:$ $\ket{\fii\phi}\ \mapsto\ U\ket{\fii\phi}$.
A \mt\ will be completed once a {\it pointer} observable of the
probe has been registered. Hence there should be a complete
collection $Q_k$ of projection operators in the probe's
Hilbert space (such that $\sum_kQ_k=I$) whose expectation values
in the state $U\ket{\fii\phi}$ give the probabilities for
the occurrence of the outcomes $k$. This {\it \mt\ scheme},
constituted by a probe, a unitary coupling, and a pointer, 
qualifies as a \mt\ of some observable of the object in the
sense that the pointer frequency distributions can be
interpreted as probabilities in the object's Hilbert
space. Indeed, to any $Q_k$ there exists an  operator
$E_k$ associated with the object such that the following
{\it probability reproducibility condition} is satisfied:
 $$
\big\langle U{\fii\phi}\big|
Q_kU{\fii\phi}\big\rangle\ =\ \ip\fii{E_k\fii}.\tag 5
$$
 for all states $\fii$.  The completeness condition $\sum_kQ_k=I$
is inherited by the $E_k$ so that $\sum_kE_k=I$. Further, all
$E_k$ are positive operators since the expectation values in
(5) are non-negative.  The map $k\mapsto E_k$
thus constitutes a positive operator valued
measure, which represents the \me{}d \ob. If the $E_k$ happen to be
projections, they define an observable in the ordinary sense, which may
be called a {\it sharp \ob}; otherwise one is dealing with an {\it
unsharp \ob}. [An
account of the general representation of observables
as operator valued measures can be found in Ref.\ 18.]

In order to distinguish
between two non-orthogonal states $\fii,\psi$ of the object,
there needs to be a projection operator $Q$ in the probe's
Hilbert space such that the corresponding probability is 1
for $\fii$ and 0 for $\psi$: thus,
 $$
QU\ket{\fii\phi}=U\ket{\fii\phi}\quad\text{ and }\quad
QU\ket{\psi\phi}=0.\tag 6
$$
It follows that
 $$
\ip{\fii\phi}{\fii\phi}\ =\ 
\big\langle U{\fii\phi}\big| U{\psi\phi}\big\rangle\ =\ 
\big\langle QU{\fii\phi}\big| U{\psi\phi}\big\rangle\ =\ 
\big\langle U{\fii\phi}\big| QU{\psi\phi}\big\rangle\ =\ 0.
\tag 7
$$
 The first equality is due to the unitarity of $U$, and we
conclude again that the state discrimination works only for
orthogonal state pairs. 

This argument is equivalent to the preceding one: if $E$ is
the effect associated with $Q$ via (5), then (6) yields (3)
and vice versa.

\subheading{4.3} {\it Why  one cannot perform \mt{}s with {\it no} state
changes.} If one could, one would be able to repeat the same
\mt\ and obtain the {\it statistics} of the \me{}d \ob\ by
manipulating a {\it single} system.
Assume a \mt\ were to leave unchanged all states of the object system.
This is to say that the expectation values of all
observables $B$ of the object would remain unchanged, irrespective
of what the \mt\ outcome was. Hence the conditional
expectation values $\langle B\rangle_k$ would have to coincide with
the original ones, i.e.,
 $$
\langle B\rangle_k\ \equiv\ \frac{\langle U\fii\phi|BQ_k|U\fii\phi\rangle}
{\langle U\fii\phi|Q_k|U\fii\phi\rangle}\ =\
\ip\fii{B\fii}\tag 8
$$
 for all states $\fii$, all $B$, all $k$. Since the
right-hand-side and the numerator of the left-hand-side are
sesquilinear functionals of $\fii$, it follows that the 
expression in the denominator must be independent of $\fii$,
 $$
\langle U\fii\phi|Q_k|U\fii\phi\rangle\ =\ \lambda_k.\tag 9
$$
 According to Eq.\ (5) the \me{}d \ob\ is
represented by the operators $E_k$, which, as a consequence
of (9), are $E_k=\lambda_k I$ if (8) is to hold. But for
such a {\it trivial} \ob\ the probabilities are the same for
all states, so that the \mt\ scheme in question does provide no
information at all about the system.

This reasoning can be refined so as to rule out the
following proposal for {\it ISD} [16]. Suppose a \mt\ with
finitely many outcomes can be performed such that the changes
of states associated with the outcomes $k$ are described
by invertible maps. Then it is possible to conceive of
further \mt{}s with the same property which, if applied to
the system after the first, principal \mt, {\it could} lead to a
reversal of the state change that had occurred in the first
instance. Thus, with some nonzero probability one would have
restored the object system's initial state; and by reading
the {\it reset} \mt's outcome, one would know this and
could repeat the principal \mt\ on the same system, in the
same initial state. It would appear that one eventually 
could collect the statistics of sufficiently many \mt{}s and
infer the state of the individual system. (Un)fortunately,
the scheme just sketched must fall under the general
category of \mt\ schemes referred to in subsection 4.2 and
is therefore doomed to fail to serve its purpose. To see why
it must fail, we observe that if a \mt\ procedure 
leads to no state change for one of its  outcomes $k$, say,
then in view of Eqs.\ (8), (9) the
corresponding effect $E_k$ is constant; therefore the probability
for that outcome does not depend on the object's initial
state. In other words, if in a sequence of \mt{}s there is a
successful {\it reset} event, all previous information about
the object state is lost.

\subheading{4.4} {\it Why the EPR-signaling scheme doesn't
work.} We consider the 2-letter alphabet version of the
signaling scheme where {\it Bob} measures either $s_x$ or
$s_y$. Then {\it Abner} either receives  one of the states
$\ket{\pm,x}$ or one of $\ket{\pm,y}$. {\it Abner} may
measure any \ob\ of particle $\Cal A$ which might allow him
to infer, at least with some probability, in which of the
two sets the state of $\Cal A$ is. Hence there should be an
outcome, represented by a positive operator $E$, whose
occurrence {\it Abner} would interpret as indicating that
the state was one of $\ket{\pm,x}$. Similarly, the
occurrence of any other outcome, the totality of which are
represented by $I-E$, would be interpreted as indicating
that the state was one of $\ket{\pm,y}$.
 Assuming equal {\it apriori} probabilities for the two sets of
states, the total probability for a correct inference is
 $$\align
 {\ts12}&\Big(\ts12\langle{+,x}|E|{+,x}\rangle\,+\,
\ts12\langle{-,x}|E|{-,x}\rangle\Big)\\ &\text{\hskip 2cm} +\,
 {\ts12}\Big(\ts12\langle{+,y}|(I-E)|{+,y}\rangle\,+\,
\ts12\langle{-,y}|(I-E)|{-,y}\rangle\Big)\\
&=\ \ts12\,\text{Trace}\left[E\cdot
\ts12\Big(\kb{+,x}{+,x}+\kb{-,x}{-,x}\Big)\right]\\
&\ \ +\,
\ts12\,\text{Trace}\left[(I-E)\cdot
\ts12\Big(\kb{+,y}{+,y}+\kb{-,y}{-,y}\Big)\right]\\
&=\ \ts12\,\text{Trace}\left[E\cdot\ts12I\right]\,+\,
\ts12\,\text{Trace}\left[(I-E)\cdot\ts12I\right]\ =\ 
\ts12.\tag 10
\endalign
$$
 Thus correct and wrong inferences are equally likely; there
is no way of telling what observable was \me{}d by the
distant observer. It would not even help if {\it Abner} and
{\it Bob} agreed that every bit of a message would be sent
in a large number of copies: if {\it Bob} had sent the $x$
letter, {\it Abner} would receive a finite ensemble of
particles which were (roughly) equally distributed
over the states $\ket{+,x}$, $\ket{-,x}$; and this would not
be distinguishable from the corresponding ensemble emerging
from sending the $y$ letter. Thus the message is blurred
as soon as one tries to employ statistical procedures.

This fact has provoked the proposal that a particle
incoming at the receiver's end should be subjected to an
amplification process by which hopefully its state could be
{\it cloned} and determined from the ensemble of copies. In
other words, the question is whether there exists some form
of amplification procedure, followed by a measurement
performed on the amplified system, which would allow {\it
Abner} to infer with certainty whether that individual
system was in one of the states $\ket{+,x}$, $\ket{+,y}$, say.
However, the process thus described constitutes a \mt\
procedure performed on these non-orthogonal states, so that
the no-go arguments of the preceding subsections immediately apply
to this situation.

\subheading{5.\ Optimal state discriminations}

\noindent
Being unable to achieve perfect state inferences, then
how well can one distinguish between two non-or\-thogonal
states [12-15]? Given one of the
non-orthogonal states $\fii,\psi$, with equal {\it apriori}
probabilities, one may ask which \mt\ would maximize the
probability of correct inferences.  Thus, again, there ought
to be an outcome, represented by a positive operator $E$,
whose occurrence is more likely in the state $\fii$ than in
$\psi$, so that one would infer that the state was
$\fii$. Accordingly, the probability for an outcome
corresponding to $I-E$ is more likely if the state was
$\psi$ rather than $\fii$. The total probability of a
correct inference is
 $$\align
 p\ &=\ \frac 12\ip{\fii}{E\fii}\,+\,\frac
12\ip{\psi}{(I-E)\psi}\\
 &=\ 
\frac
12\left(1+\text{Trace}\Big[E\cdot\big(\kb\fii\fii-\kb\psi\psi\big)
\Big]\right).\tag 11
\endalign
$$
 It is obvious that this expression is maximal if $E$ is
the spectral projection associated with the positive
eigenvalue of the operator $\kb\fii\fii-\kb\psi\psi$.  The
eigenvalues are easily determined to be
$\pm\sqrt{1-|\ip\fii\psi|^2}$, so that the maximal available
probability is
 $$
p_{\text{max}}\ =\ \frac
12\Big[1+\sqrt{1-|\ip\fii\psi|^2}\Big].\tag 12
$$
 As an example, in the case of a \soh\ system and with
$\fii=\ket{+,x}$, $\psi=\ket{+,y}$ one finds
$p_{\text{max}}=\frac 14\big[2+\sqrt 2\big]\approx 0.85$.

There are other ways of formulating an optimization problem.
For instance, one may conceive of a \mt\ scheme which allows
one to infer $\fii$ and $\psi$ with certainty from some
distinct outcomes, but at the price that there are some
further outcomes from which no unique inferences are
possible. Thus there should be (at least) three outcomes, represented
by the probe projections $Q_1,Q_2,Q_3$ (where
$Q_1+Q_2+Q_3=I$), such that
 $$
Q_2U\fii\phi=0,\qquad  Q_1U\psi\phi=0.\tag 13
$$
 Under these circumstances it is evident that the state must
have been $\fii$, resp.\ $\psi$, if the outcome is 1, resp.\
2. According to Eq.\ (4) there are three positive operators
$E_1,E_2,E_3$ associated with the object system, which allow
one to rewrite Eqs.\ (13) as
 $$
E_2\fii=0,\qquad E_1\psi=0.\tag 14
$$
 We wish to maximize the  probability of correct inferences
(assuming again equal {\it apriori} probabilities),
 $$
p\ =\ \frac 12\ip\fii{E_1\fii}\,+\,\frac 12\ip\psi{E_2\psi}.
\tag 15
$$
 There is a constraint: the residual operator
$E_3=I-E_1-E_2$ must be positive,
 $$
\ip\xi{(I-E_1-E_2)\xi}\ \ge\ 0 \qquad\text{for all states\ }
\xi.\tag 16
$$

For the sake of simplicity we consider only the case where
the object's Hilbert space is two-dimensional (e.g., a \soh\
system). Then the conditions (14) imply that $E_1,E_2$ are
of the form
 $$
E_1\ =\ e_1\big(1-\kb\psi\psi\big),\qquad
E_2\ =\ e_2\big(1-\kb\fii\fii\big).\tag 17
$$
 Note that the positivity condition (16) forbids $e_1,e_2$
to assume their maximal value 1; this is to say that the
operators $E_1,E_2,E_3$ cannot be projections. In other
words, they constitute an {\it unsharp observable} of the
object [18]. Inserting (17) into (15) yields
 $$
p\ =\ \frac
12\big(e_1+e_2\big)\,\big(1-|\ip\fii\psi|^2\big),\tag 18
$$
 and the positivity constraint (16) is found to be
 $$
0\ \le\ 1-\big(e_2+e_2\big)+
e_1e_2\big(1-|\ip\fii\psi|^2\big),\tag 19
$$
 Noting that the probability $p$ increases with increasing
$e_1,e_2$, we solve (19) for $e_1$, 
 $$
e_1\le \frac{1-e_2}{1-e_2w},\quad 
w\equiv{\big(1-|\ip\fii\psi|^2\big)},\tag 20
$$
and insert this into (18):
 $$
p\ \le\ \frac 12 f(e_2)w,\quad
f(e_2)=\frac{1-e_2^2w}{1-e_2w}.\tag 21
$$
 The function $f(e_2)$ is maximal at $e_2=\frac
1w\big[1-\sqrt{1-w}\big]$, for which the maximal value of
$e_1$ according to (20) is $e_1=e_2$. Therefore the maximal
value of $p$ is
 $$
p_{\text{max}}\ =\ 1-\big|\ip\fii\psi\big|\qquad
\text{at }\ e_1=e_2=\frac 1{1+\big|\ip\fii\psi\big|}.\tag 22
$$
 
This reproduces the result of Jaeger and Shimony [15] and
underlines it with the observation that optimal state
inference in the present sense and case is achieved with a
\mt\ of an unsharp rather than sharp \ob.

\subheading{6.\ Conclusion} 

\noindent
 {\it Is the quantum state (an) observable? } In physics one
wants to find out about the {\it state} of a system; and all
one can do is to perform \mts\ of some {\it observables}.
There is no way to obtain a unique determination of an
unknown state from a {\it single} \mt\ outcome. The reliability of the
state discrimination decreases with increasing
similarity between the two states under consideration,
measured in terms of their inner product.

It may be worth recalling that
there are \mt\ procedures of a single observable 
which give complete {\it statistical}
information about the state of a system [18]. Such an {\it
informationally complete} \ob\ is necessarily unsharp.
In fact for any sharp observable $A$, the
vectors $\psi$ and $e^{if(A)}\psi$ represent in general different
states [if $\psi$ is not an eigenvector of $f(A)$]; but
these states have the same $A$ distributions.

We conclude that the quantum state is not an observable but not
unobservable.

\subheading{Acknowledgement}

\noindent
 This work was supported by the Alexander von Humboldt
Foundation. Hospitality of the Department of Physics,
Harvard University, where part of this work was carried out,
is gratefully acknowledged. My sincere thanks go to Sidney Coleman for
inspiring discussions on the subject of this paper.

\subheading{References and Notes}

{\advance\baselineskip by -6pt
{

\item{1.\ } N.\ Herbert, {\it Found.\ Phys.} {\bf 12}, 1171 (1982).

\item{2.\ } D.\ Dieks, {\it Phys.\ Lett.} {\bf 92A}, 271
(1982).

\item{3.\ } P.\ Eberhard, {\it Nuovo Cim.} {\bf 46B}, 392
(1978).

\item{4.\ } G.C.\ Ghirardi, A.\ Rimini,\ T.\ Weber, 
  {\it Lett.\ Nuovo Cim.} {\bf 27}, 293 (1980).

\item{5.\ } R.J.\ Glauber, in {\it New Techniques and Ideas
in Quantum Measurement Theory}, ed.\ D.M.\ Greenberger, The
New York Academy of Sciences, New York, 1986, pp.\ 336-372,
esp.\ p.\ 362ff.

\item{6.\ } L.\ Mandel, {\it Nature} {\bf 304}, 188 (1983).

\item{7.\ } W.K.\ Wootters, W.H.\ Zurek, {\it Nature} {\bf
299}, 802 (1982).

\item{8.\ } The question of the {\it peaceful coexistence} of
quantum mechanics and relativity has been reviewed by
A,\ Shimony, in S.\ Kamefuchi et.\ al.\
(eds.), {\it Foundations of Quantum Mechanics in the Light
of New Technolgy}, Tokyo: The Physical Society of Japan, 1984.

\item{9.\ } A formulation of the no-signaling proof that
comprises general amplification and cloning procedures is
given in H.\ Scherer, P.\ Busch, {\it Phys.\ Rev.} {\bf
47}, 1647 (1993).

\item{10.} A.\ Peres, {\it Quantum Theory: Concepts and
Methods}, Kluwer Academic Publishers, Dordrecht, 1993,
Chapter 9.

\item{11.} M.G.\ Alford, S.\ Coleman, J.\ March-Russell,
{\it Nuclear Physics} {\bf B351}, 735 (1991).

\item{12.} I.D.\ Ivanovic, {\it Physics Letters A} {\bf
123}, 257 (1987).

\item{13.} D.\ Dieks, {\it Physics Letters A} {\bf 126}, 303
(1988).  

\item{14.} A.\ Peres, {\it Physics Letters A} {\bf 128}, 19
(1988). 

\item{15.} G.\ Jaeger, A.\ Shimony, {\it Phys.\ Lett.\ A}
{\bf 197}, 83 (1995).

\item{16.} A.\ Royer, {\it Phys.\ Rev.\ Lett.} {\bf 73}, 913
(1994); Erratum: ibid., {\bf 74}, 1040 (1995).

\item{17.}
 We prove the following mathematical fact which will be used
in several instances throughout this paper: for an operator
$E$ with $0\le \ip\xi{E\xi}\le 1$ for all states $\xi$, the
relation $\ip\fii{E\fii}=1$, resp.\ $\ip\psi{E\psi}=0$ (for
states $\fii,\psi$) is equivalent to $E\fii=\fii$, resp.\
$E\psi=0$.  The latter equations are obviously sufficient
for the former. Their necessity follows from the
observation that $\ip\xi{F\xi}=\big\langle{\sqrt F\xi}\big|{\sqrt
F\xi}\big\rangle=\parallel{\sqrt F\xi}\parallel^2$, where $F$ is
either $E$ or $I-E$.

\item{18.} P.\ Busch, M.\ Grabowski, P.\ Lahti, {\it
Operational Quantum Physics}, Springer-Verlag, Berlin, 1995.

}}
\bye